# Towards Auditability Requirements Specification Using an Agent-Based Approach

Denis J. S. de Albuquerque[1], Vanessa Tavares Nunes[1], Claudia Cappelli[2] and Célia Ghedini Ralha[1]

[1]Department of Computer Science, University of Brasilia, Brazil
[2] Informatics Graduate Program, Federal University of Rio de Janeiro, Brazil

## Abstract

*Transparency is an important factor in democratic societies composed of characteristics such as accessibility, usability, informativeness, understandability and auditability. In this research we focus on auditability since it plays an important role for citizens that need to understand and audit public information. Although auditability has been a subject of discussion when designing systems, there is a lack of systematization in its specification. We propose an approach to systematically add auditability requirements specification during the goal-oriented agent-based Tropos methodology. We used the Transparency Softgoal Interdependency Graph that captures the different facets of transparency while considering their operationalization. An empirical evaluation was conducted through the design and implementation of LawDisTrA system that distributes lawsuits among judges in an appellate court. Experiments included the distribution of over 300,000 lawsuits at the Brazilian Superior Labor Court. We theorize that the presented approach for auditability provides adequate techniques to address the cross-organizational nature of transparency.*

## 1. Introduction

Organizations have been evaluated in their ability to provide auditable information as a support for trusting in their operations, performance, and results [1]. The aim is to improve people's views of processes and information to provide awareness, to reduce omission, to enable control, to facilitate research, and to increase trust. In this regard, being auditable is an important concern when designing systems that support the execution of processes and manipulation of information. Auditability has been associated as one of the characteristics that promote organizational transparency. Although transparency has been a subject of discussion to the openness of organizations, especially concerning systems design, the systematization of requirements specification is still an open problem that configures a double challenge: (1) what this concern exactly means in systems' development; and (2) how to elicit, model, and design it in a system.

First, transparency is seen as a quality issue and needs to be detailed in a more concrete perspective as the authors in [2] proposed, where transparency can be achieved by the implementation of five characteristics: accessibility, usability, in formativeness, understand ability, and auditability. These transparency characteristics were defined in a Transparency Software Interdependency Graph (SIG) that aims to define a transparency catalog to be systematically implemented in organizational processes and information. Being these characteristics orthogonal to organizations, requirements to provide auditability in systems may impact the choices on how the functionalities are to be designed and implemented.

Second, transparency characteristics such as auditability can be seen as softgoals, which are goals that do not have a clear-cut criterion for their satisfaction and are usually measured with a more positive or negative evidence of achievement. Considering the requirements specification goal-oriented agent techniques have enjoyed significant attention [3]. This approach is motivated by the need to overcome the semantic gap between systems and the organizational environment in which they operate since they adequately address the cross-organizational nature of transparency. Therefore, we propose an approach to systematically add auditability requirements specification during the goal-oriented agent-based design phase using Tropos methodology [4]. We relied on the auditability operationalization proposed in the Transparency SIG [2]. This paper is an evolution of [5] [6], where transparency characteristics related to understandability and an early perception of auditability were designed and tested in a prototype.

The proposed approach was empirically evaluated through the development of LawDisTrA (Lawsuit Distribution Transparent Agents), an agent-based system to address the automatic distribution of judicial processes (lawsuits), using real data from the Brazilian Superior Labor Court (TST). Although the lawsuit distribution automation improved transparency, it is still criticized by the lack of aspects, such as execution, auditability, and understandability[1][2].

The rest of the paper is organized as follows: In Section 2, we present the research background on organizational transparency, agent-based theory, and their integration when analyzing auditability characteristic during a goal-oriented agent-based system design; in Section 3, our approach is proposed using the Brazilian lawsuit process flow to illustrate the LawDisTrA system whose development is fully described, including the applied auditability characteristics, the requirements specification, and the architectural and implementation aspects; in Section 4, LawDisTrA is evaluated under the Brazilian lawsuits distribution at the TST; in Section 5, an analysis of the expected contributions and related work is discussed; and in Section 6, we conclude the paper and present future work.

## 2. RESEARCH BASELINE

This section details the Transparency SIG proposed in [2] whose characteristics we relied on during the auditability requirements specification. Next, we discuss agent-based systems and their close relation to transparency aims to expound why we designed and implemented LawDistrA as an agent-based system. Ultimately, we introduce goal-oriented requirements specification through the Tropos methodology [4].

### 2.1. Transparency Software Interdependency Graph

Transparency is the social value of open and public access to information [2]. Thus, transparency represents a new and important concern that developers have to deal with. It establishes a set of aspects that suggest the existence of policies, standards, procedures, and technologies.

---

[1] http://www.conjur.com.br/2014-jul-12/advogados-exigem-transparencia-relacao-processo-eletronico
[2] http://www.trt4.jus.br/portal/portal/trt4/comunicacao/noticia/info/NoticiaWindow?cod=1378302&action=2

Following this path, the Transparency SIG [2] aims to define a non-functional requirements (NFR) catalog, based on [7], that provides five groups of characteristics related to accessibility, usability, informativeness, understandability, and auditability. For each characteristic, a set of operationalization and mechanisms were proposed to orient their implementation. As an example, the characteristic of Controllability, from the Auditability (Figure 1), is defined as the ability to have rule over something. To operationalize it, the organization must implement actions to calculate estimated versus realized, perform simulation, register, start and end of activities, register participating actors, register decisions taken, define process control points, check information, and register used resources. The mechanisms to implement this operationalization can be systems, policies, rules, and others.

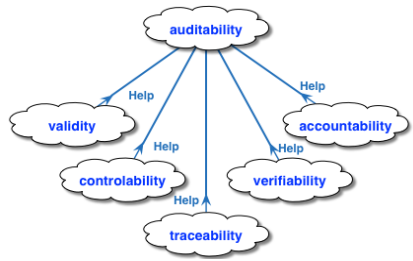

Figure 1. Auditability Transparency Group

## 2.2. Agent-based Systems

An agent-based system is described as a collection of software entities that interact in a computer network to achieve individual and organizational goals. An agent is a software entity able to perceive its environment and to act on it. The agent-based software engineering paradigm focuses on systems that take place in a dynamic and uncertain environment, can engage in social interactions and can operate within flexible organizational structures [8].

In this regard, agent-based systems provide an interesting way to simulate organizations by representing real-world problems in their natural complexity, which may help to shed some light on various kinds of social processes [9]. The design of how agents reason, act, and interact with each other and with the environment is essential to understand and define the levels of collaboration, negotiation, clarity, integrity, accountability, traceability, and other characteristics that also compose the concept of transparency.

Regarding agents' intelligence, there is a particular type of knowledge-based system cited as deductive reasoning agents [8]. Deductive reasoning agents use logic, typically defined as a set of rules, to encode a theory defining the best action to perform in a situation. The selection of an action is done via a rule engine tool that implements inferences through the use of forward and backward chaining algorithms. In this work, we used Drools[3], a free Business Rule Management System (BRMS) developed in Java language. The implementation of LawDisTrA system uses JADE[4] (Java Agent DEvelopment Framework), a middleware to agent-based systems development, compliant to FIPA[5] standards [10].

---

[3] JBoss Drools – http://www.drools.org
[4] JADE - http://jade.tilab.com/
[5] FIPA (The Foundation for Intelligent Physical Agents) is an IEEE organization that promotes standardization of technologies related to agents and their interoperability

## 2.3. Goal-oriented Requirements Specification

Among the existing goal-oriented methodologies, this research used Tropos, which is a requirements-driven methodology that seeks to support various phases of an agent-based system development [4]. Tropos models are built to capture the intentions of the stakeholders (e.g., users, owners) which is a challenging task [11]. It adopts the ISTAR (Intentional STrategic Actor Relationships) modelling framework to represent the actors' concepts (agents, positions, or roles), objectives (hardgoals and softgoals), tasks, resources and their interdependencies (dependency links, contribution, or decomposition, and decomposition and means-end) [12]. Figure 2 presents the ISTAR framework notation representing these concepts.

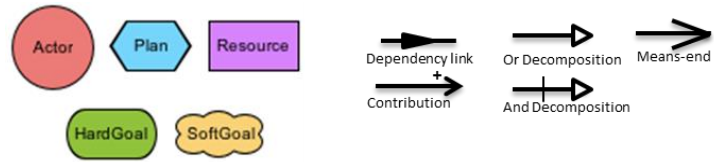

Figure 2. ISTAR Notation

Tropos also goes in line with transparency characteristics that are considered qualitative goals, i.e., softgoals (following their NFR characterization) and transversal to the system.

## 3. RELATED WORK

Research on applied technologies to electronic services has increased in the last few years. Indeed, the introduction of information and communication technologies may be a vehicle to create a culture of transparency and access to information [13].

Approaches for different purposes have been used to apply agent-based technology to electronic services. An agent-based framework is proposed for the managing of the Indian Public Food Distribution System that has been largely criticized due to its lack of transparency. It enhances the capacity of understanding the impacts on decisions about the supply chain, although transparency as a system requirement was not formally applied [14]. Carneiro et al. [15] proposed an agent-oriented architecture to increase effectiveness and awareness of dispute resolution of court processes. Authors expand the amount of meaningful information and possible outcomes available for the parties using case-based reasoning techniques. Transparency is considered a fundamental characteristic but treated as an abstract quality of the solution.

Müller et al. [16] discuss standard approaches for modelling descriptions of agent-based systems and enforce the need to provide transparent models as a requirement. The authors do not define transparency and focus on the discussion of Web Ontology Language (OWL) as a modelling language used to improve transparency of agent-based systems' formal descriptions. Serrano and Leite [17] proposed an approach to capture requirements patterns through argumentation by identifying, on argumentation graphs, NFR operationalization of software transparency characteristics[6]. The research focuses on applying transparency to the software and the software development process, while our research focuses on applying transparency in the process and information through the use of the software that supports it.

Hosseini et al. [18] proposed four reference models for systems transparency requirements. They aim to enable requirements engineers and software analysts to better manage the stakeholder's transparency requirements through a holistic conceptual baseline. The reference models cover transparency actor, meaningfulness, usefulness, and information quality. Those are important viewpoints to deal with to implement transparency in systems, but they do not consider a broader and complete set of viewpoints that organizational transparency relates to. Nevertheless, authors worked on an information system focused viewpoint operationalization, deepening its understanding and way of using it at the design time. We believed this approach is correlated with our research and may help to establish a systematic way to think about each transparency characteristic more deeply.

---

[6] http://transparencia.inf.puc-rio.br/wiki/index.php/Cat%C3%A1logo_Transpar%C3%AAncia

Therefore, systematic approaches that treat transparency through the use of systems are still in their initial stage. Our approach enhances systems development using an agent-oriented paradigm. In this work we present a systematic way of designing transparency characteristics as systems' requirements illustrated on a specific scenario.

## 4. AUDITABILITY REQUIREMENTS SPECIFICATION: A SCENARIO THROUGH LAWDISTRA

In Section 4.1, we present our approach to enhance Tropos to systematically add auditability requirements analysis and design. In Sections 4.2 to 4.6, we present LawDisTrA including the design and development conceived through the five phases of Tropos methodology [4].

### 4.1. Auditability Analysis and Design

Tropos methodology has five phases: early requirements, late requirements, architectural design, detailed design, and implementation [4]. This research proposes an additional and first step of the late requirements phase, where transparency analysis is performed to evaluate the agent-based system's early requirements and to guide the late requirements definition. The idea behind this proposition resides in the fact that transparency analysis may (i) demand a change in how some of the elicited agents' plans must be performed; (ii) define the systems' NFRs; and/or (iii) be related to the agent-based system itself.

During this step, we propose that each characteristic of the Transparency SIG (Figure 1) has to be analysed and prioritized according to the organizations' transparency needs. For each one of the 28 transparency characteristics (leaves) [2], their operationalization has to be analyzed, and domain-specific softgoals must be proposed by the requirements analysts. These softgoals are analysed to define late requirements, architectural demands, and NFRs. To experiment the proposed approach focusing on the auditability aspect, we designed and developed LawDisTrA, a system to address the automatic distribution of judicial processes (lawsuits) in Brazil. The approach is fully described in this scenario.

### 4.2. LawDisTrA Early Requirements

In the Early Requirements phase, the problem and the organizational context were understood, and three agent types were identified based on the organizational structure: Protocol Agent (PA), Magistrate Agent (MA), and Distribution Agent (DA). PA represents the persons or departments responsible for preparing lawsuits for distribution. MA represents the Magistrates responsible for their judgments. Magistrates compose Judicial Bodies (JBs). DA represents the persons or departments responsible for performing the distribution of lawsuits to JBs and MAs.

#### 4.2.1. Lawsuit Distribution Scenario

The lawsuit process flow orchestration is an instrument that aims to ensure fundamental rights to citizens and to protect the legal order, being subjected to law principles, such as legality, efficiency, and natural justice. The lawsuit consists of case files - constituent parts of a process, such as petitions, certificates, terms, among others. The process' life cycle starts when lawsuits are submitted to the TST, which is to try lawsuits arising from labor relations. They are treated and distributed to the available magistrates that judge sentences to be published.

The lawsuit distribution defines for each lawsuit, under the Civil Procedure Code and rules established by the judicial power, the judging organ, and Minister. These rules exist to ensure impartiality and fairness since they seek to preserve the free distribution rule and the principles of natural justice.

The principles of natural justice, generally known as the duty to act fairly, are highlighted in Art. 5 of the Brazilian Federal Constitution[7]. Their scope is to guarantee independence and impartiality of the judging organ, by ensuring the administrative action befits the principles of equality, legality, impersonality, morality, publicity, and efficiency, provided in Art. 37 of the Brazilian Federal Constitution. The free distribution rule states that the free drawing of lawsuits (among the 27 TST Ministers) is the default rule when there is more than one competent judge to decide on the case, respecting interdependencies to promote uniform and fair judgments in connected cases; legal impediments (according to Art. 144 of the Civil Procedure Code) that forbid a magistrate to act on it; and suspicious situations which are subjective, although Art. 145 of the Civil Process Code define situations that must be analysed. When there are suspicions or impediments, the process is to be redistributed. All these issues and how they affect the lawsuit distribution should be common knowledge among society. Therefore, the Information and Communications Technology (ICT) supporting the lawsuit process flow represents a major concern in attaining these requirements. We can also bring to this context what the authors in [19] discussed about how it is important to detail requirements analysis and design when in regulated environments where characteristics such as consistency, unambiguity, verifiability and traceability are a major concern.

Authors in [20] indicate that the participating lawyers recognized that ICT support enhanced transparency in general, although they are not unanimous in trusting completely the information and process managed by the electronic government solution. In Sections 3.3 to 3.6, we describe LawDisTrA design and implementation, considering transparency in agent-based systems.

## 4.3. LawDisTrA Late Requirements

In the Late Requirements phase, the first step was to elicit the auditability requirements that were to be analysed and designed during the late requirements and the following phases. Our proposal for auditability requirements specification is fully described in Section 3.3.1. Thereafter, and considering all auditability requirements elicited, action plans were specified during this phase for each of the three agent types (i.e., PA, MA, DA). As depicted in Figure 3, considering the DA type, each agent must achieve hardgoals (in green) and softgoals (in beige) and manipulate data assets (in purple) by performing action plans (in blue). The DA's main goal is to perform the lawsuit distribution. Therefore, it needs to interact with the Distribution Database (where the lawsuit processing is stored) and the other agents to identify available JBs; identify available Magistrates, along with their impediments and JBs' composition; identify lawsuits waiting for distribution; apply defined distribution rules; and inform the distribution so that the PA can update the database of court proceedings.

### 4.3.1. Auditability Design

We propose that, as a first step in the late requirements phase, the solution is to focus on the auditability characteristic (and later on the other transparency characteristics) to systematically provoke requirements analysts to think of what is important (the softgoals in grey in Figure 4) to elicit when considering auditability for this specific system (colored softgoals in Figure 4).

---

[7] http://www.planalto.gov.br/ccivil_03/Constituicao/Constituicao.htm

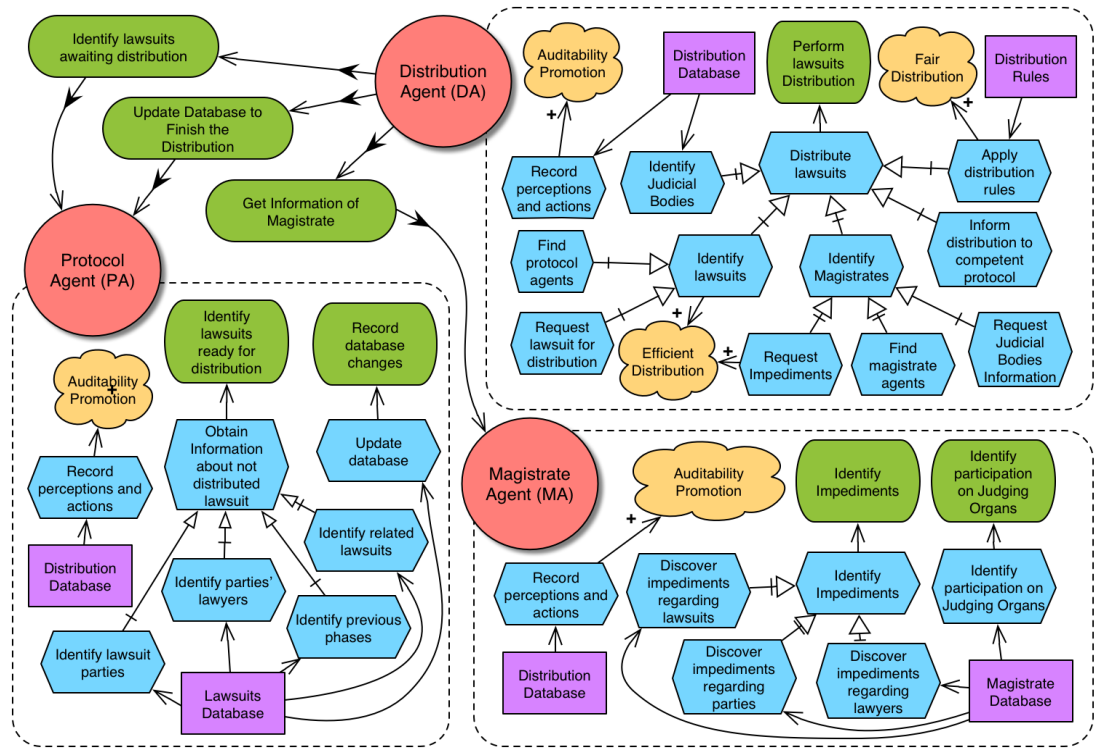

Figure 3. LawDisTrA Late Requirements model

The analysis and design of auditability characteristics in an agent-based system are shown under the LawDisTra scenario. As presented in Figure 1, there are five branches of the auditability group. For each of them, a set of LawDisTra operationalization (see Figure 4) was proposed based on the generic operationalization (in grey), proposed by the Transparency SIG [2]. It is possible to observe that some operationalizations help to implement more than one characteristic of auditability (marked with repeated colors). The operationalization was written following the Transparency SIG format.

As an example, analysing the traceability operationalization, the following actions are demanded:

- Each agent must have its behaviors defined (in red): In the detailed design phase, agents' behaviors need to be modelled, including their possible actions and communication interfaces among agents, to attend the Transparency SIG operationalization "Identify software requirements x activities" (in grey).
- Each agent must register date, actors involved, location, process status, and justification of each information change it is responsible for (in white): The plan "Record perception and actions" (Figure 3, presented in all agent's boxes) must include the registration of the previously defined properties of each information change, and the information must be shown to interested parties. This mechanism is to attend the operationalization "Identify the context of changes," "Identify when changes are performed," "Identify the location of change," and "Identify responsible for changes" (all in grey).
- The agents must identify and register a change and its reason in the default distribution process (in white): The plan "Record perception and actions" (Figure 3, presented in all agent's boxes) must include the registration of each change in the distribution process. The information must be shown to interested parties. It is to attend the operationalization "Identify changes reasons" and "Identify the process changes."

- Each agent must document in which distribution instance the information was manipulated (in green): The plan "Distribute lawsuits" (Figure 3, in DA's box) must register the distribution process instance identifier. The information must be shown to interested parties. It is to attend the operationalization "Identify information x process instances."
- The agents must identify dependencies among process instances related to lawsuits and magistrates (in white): The plan "Apply distribution rules" (Figure 3, in DA's box) must consider dependencies among process instances concerning their respective set of lawsuits. Therefore, there must be business rules to register this identification and to correlate process instances and show this information to the parties involved. It is to attend the operationalization "Identify dependencies among processes."
- The agents must be aware of and register the agents (and inputs) that activate it and the agents (and outputs) it activates (in blue): Each agent has a plan "Record perception and actions" (Figure 3, presented in all agent's boxes) that must register all architectural and detailed design that models agents' behaviors and interaction relations. The information must be shown to interested parties. It is to attend the operationalization "Identify predecessor activities" and "Identify successor activities."

It is possible to observe that some operationalizations are related to the software development process and others are related to the implementation of functional requirements and visualization mechanisms. For example:

- The operationalization "Each agent must register date, the actor involved, location, process status and justification of each information change it is responsible for" requires that the design of the plan "Record perceptions and actions" include this information (see DA plan in Figure 3). Therefore, during the Detailed Design phase, the data model must consider them.
- The operationalization "The agents must be aware of and register the agents (and inputs) that activate it and the agents (and outputs) it activates" requires that during the Architectural Design phase the communication architecture between agents must be schematized accordingly.

In this scenario, there was no need to change the late requirements plans by adding new ones, but rather to improve what some of the plans may realize and how the following phases should be designed and constructed. Sections 3.4 to 3.6 present how these requirements were considered in the LawDisTra design and implementation.

### 4.4. LawDisTrA Architecture Design

In the Architecture Design phase, the interconnections between LawDisTrA agents and JADE framework special agents were identified.

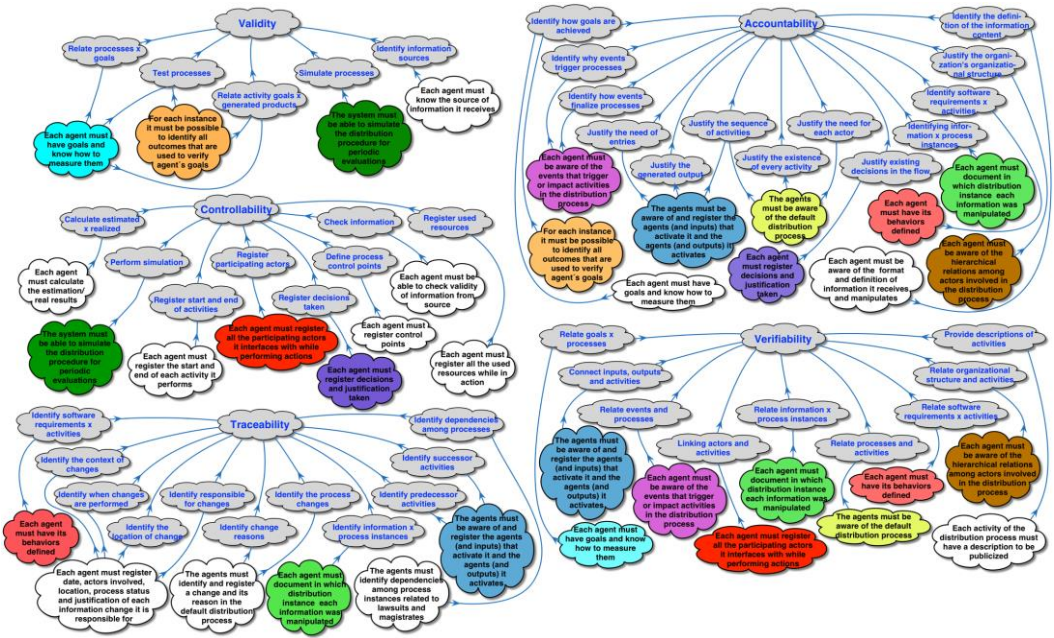

Figure 4. Auditability operationalization

### 4.4.1. The Architecture

The architecture includes two special agents (agent management system - AMS and directory facilitator - DF) that are automatically activated with JADE middleware [10]. The AMS controls the agent platform (white pages), dealing with the creation, completion, and other stages of the agents' life cycles. The DF is an agent that provides a directory service (yellow pages) and discloses to all agents, the available agents, and services. Figure 5 depicts LawDisTra architecture, which is designed to preserve as much resemblance to the TST environment. The arrows in the diagram indicate the direction of data flow between components. For example, the "Distribution Auditor" component consumes data from the "Electronic Judicial Processes Database" and the "Distribution Database.".

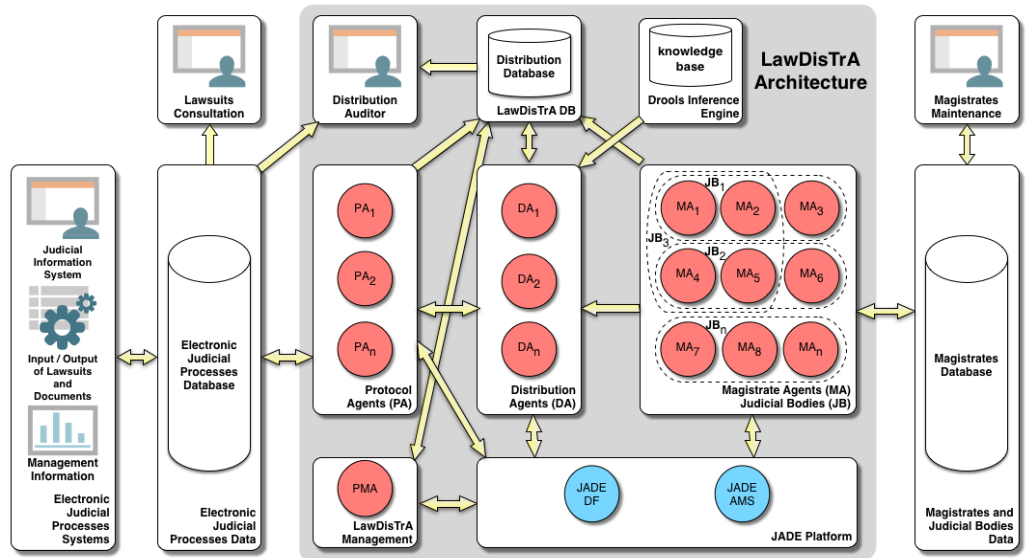

Figure 5. LawDisTrA Architecture

This architecture layout demonstrates that:

- There may be several PAs (PA1, PA2, ..., PAn), each one specialized in treating specific types of lawsuits.

    o PA must identify in the "Electronic Judicial Processes Database" which lawsuits require distribution, obtaining the necessary information to carry it out. They must have the intelligence to identify, for example, when a lawsuit is related to others, so they can be distributed to the same MA;
    o Information obtained by PAs will be passed on to DAs on request. PAs may request the DAs to initialize distribution procedures;
    o PAs update the "Electronic Judicial Processes Database" with data from lawsuits movement processing (across departments) after distribution information received from DAs;
    o PAs update the "Distribution Database" to record their perceptions and actions, allowing these data to be used in any queries by the "Distribution Auditor."

- There are several MAs (MA1, MA2, ..., MAn), each one representing one magistrate in the court, and JBs (JB1, JB2, ... JBn) that are logical groupings of MAs. JBs define the powers of magistrates to judge lawsuit cases. Each MA can be a member of many JB.

    o MAs identify in the "Magistrates Database" the JBs of which they are members and their respective impediments. This information will be passed on to DAs upon request;
    o MAs may have the intelligence to autonomously identify new impediments regarding a magistrate from other data sources and update the "Magistrates Database." As an example, an MA could find out that a magistrate is related to one person involved in the lawsuit;
    o MAs also update the "Distribution Database" to record their perceptions and actions.

- Several DAs (DA1, DA2, ..., DAn) can be used to divide the processing load or to focus on a specialized distribution where a set of DAs treat specific types of lawsuits. For example, a DA could be employed for each available PA.

    o DAs obtain the necessary information for distribution through interaction with PAs and MAs. They do not directly access the "Electronic Judicial Processes Database" and the "Magistrates Database," to avoid coupling and increase the LawDisTrA architecture flexibility;
    o Lawsuit distribution can occur automatically, with DAs performing their tasks in a continuous or scheduled basis or by direct interaction with users;
    o DAs record in the "Distribution Database" the necessary information for distribution, as well as their perceptions and actions. This information can be used both by other DAs while assisting the ongoing distributions or by the "Distribution Auditor" for transparency purposes;
    o Distribution rules used by the DAs are specified separately in the Drools knowledge base whose inference engine, combined with DAs' perceptions, indicates the rules to use in lawsuit distribution. DAs' perceptions, built from their interaction with other PAs and MAs, are the facts of the knowledge base.

International Journal of Software Engineering & Applications (IJSEA), Vol.11, No.3, May 2020

- PMA (Platform Manager Agent) was defined to assist LawDisTrA management by the user who can activate or deactivate PAs, MAs, and DAs, as well as monitor the system operation.
    - PMA carries out its actions through interaction with JADE platform agents;
    - PMA queries the "Distribution Database" to obtain LawDisTrA settings, identifying which are the available protocols, judges, and distributors;
    - PMA also records its perceptions and actions in the "Distribution Database." They can be used to track when the system and the agents are loaded or unloaded, and other information concerning the LawDisTrA operation.
- The "Lawsuits Consultation" component is a publicly available judicial information query system and provided by the TST[8].
- The "Distribution Auditor" is designed to promote transparency of the information in the "Distribution Database" so that one can audit each one of the distributions. It also uses information extracted from the "Electronic Judicial Processes Database" to present the lawsuit to the user with detailed information. It leverages the Lawsuit Consultation component by adding more details and allowing the tracking of LawDisTrA agents.
- The "Magistrates Maintenance" component is responsible for keeping the information related to the Court's judges in the "Magistrates Database."
- The "Electronic Judicial Processes Systems" and "Electronic Judicial Processes Database" in the TST are managed and worked by a set of systems as follows:

    - "Judicial Information System" that performs the content treatment and the movement between the various administrative units that act in the lawsuit flow from its assessment to the trial and publication of the decision;
    - "Input/Output of Lawsuits and Documents" that handles lawsuits coming/going from/to a lower court instance or originating from the Federal Court itself; and
    - "Management Information" that uses lawsuit data for statistical processing and preparation of management reports to support decisions.

## 4.5. LawDisTrA Detailed Design

During the Detailed Design phase, the system was modeled in deeper detail. At the capability level, agents and their behaviors were modeled according to JADE constraints and transparency requirements definition using Unified Modeling Language (UML) class diagrams. The data level was modeled through the design of a conceptual data model (via entity-relationship diagram) that represents a view of the TST database, as presented in Figure 6.

---

[8] http://aplicacao4.tst.jus.br/consultaProcessual

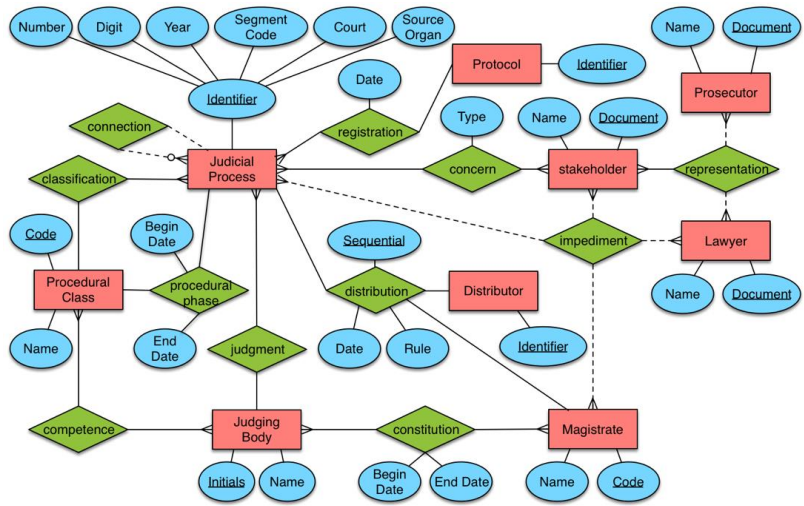

Figure 6. Entity-relationship diagram of LawDisTrA database

The Judicial Process is assessed by a Protocol entity that represents a department of the TST organization. The Judicial Process is judged by Judging Bodies that have the competence to judge processes according to their Procedural Class. A Judicial Body is composed of Magistrates who will be responsible for conducting the judgment of the processes distributed to them by a Distributor (department or person of the organization responsible for the distribution). The Magistrate may have impediments related to lawyers, specific court cases, or parties to the proceedings. The parties are persons or organizations interested in certain Judicial Processes which are represented by Lawyers or Prosecutors (where the represented is a government entity). Each entity has its attributes illustrated by the ellipses in blue.

### 4.6. LawDisTrA Implementation

During the Implementation phase, the codification was developed following the requirements and architectural definitions. It is important to notice that the capability level goes beyond the distribution codification, including interface development that has a great impact when implementing transparency. At the data level, the logical data model was defined in the Database Management System (DBMS). A knowledge level was implemented using Drools to separate the business rules from the application logic. Thus, it is possible to promote transparency in the implementation of business rules, as opposed to using stored procedures as currently done in the TST. A forward-chaining algorithm was used, so the agents can act while obtaining the necessary information from the environment.

#### 4.6.1. The Implementation description

LawDisTrA was implemented in Java, apart from the "Distribution Auditor" which was implemented using Dart[9] (on SDK version 1.16). Source codes and all models are available at https://gitlab.com/InfoKnow/Transparency/LawDisTrA. JADE[10] version 4.4.0 was used as an agent-based middleware.

Four distribution rules were implemented using Drools (src/br/unb/sma/rules/Distribution.drl): Rule 1: distribution by the dependency of lawsuits that are related to other lawsuits already distributed; Rule 2: distribution by prevention of existing lawsuits that are in new phases and must be distributed to the same Judicial Body and Magistrate of the previous phase; Rule 3: distribution of embargoes due to divergent decisions among Judicial Bodies; and Rule 4: ordinary distribution by drawing of Judicial Bodies and Magistrates (general distribution).

---
[9] https://www.dartlang.org/
[10] http://jade.tilab.com/

LawDisTrA interface was implemented using Java Swing. For each agent, a GUI interface was created to monitor its execution. The interface for the PMA agent was developed to allow the manager user to monitor and control (activate and deactivate) PAs, MAs, and DAs.

The "Distribution Auditor" was implemented as a web interface (with some similarities to the TST system) allowing access to lawsuits and distribution data to provide transparency to users. Figure 7 illustrates a query on the lawsuit AIRR 3128-70.2012.5.18.102. We can observe the Processual class (Classe), the initial date of the current phase (Início da fase atual), and to which Judicial Body (Órgão Judicante) and Magistrate (Relator) the lawsuit was distributed. It is possible to see the interested parties (Partes) (in blurred lines due to privacy issues). Note that the labels appear in Portuguese since they were built for use in a Brazilian Court.

To improve transparency, the interface also presents the following details regarding distribution (Distribuição): distribution number (Número); date of distribution (Data); rule activated in the distribution (Regra); distribution result (Resultado); drawing details (Detalhes do sorteio), like competent and available Judicial Bodies (O.J. competentes), selected Judicial Body (O.J. sorteado), selected Judicial Body composition of Magistrates (Composição do O.J.), impeded Magistrates (Magistrados impedidos), and selected Magistrate (Magistrado selecionado); and reasons for the impeded Magistrates (Impedimentos para MMCP).

Pressing the button "Rastrear Agentes" the system loads the tracking interface, as depicted in Figure 7, which allows the user to check the agents' actions related to the distribution and that were registered in the Distribution Database. Figure 8 shows 12 out of 93 records that were made for this specific lawsuit distribution.

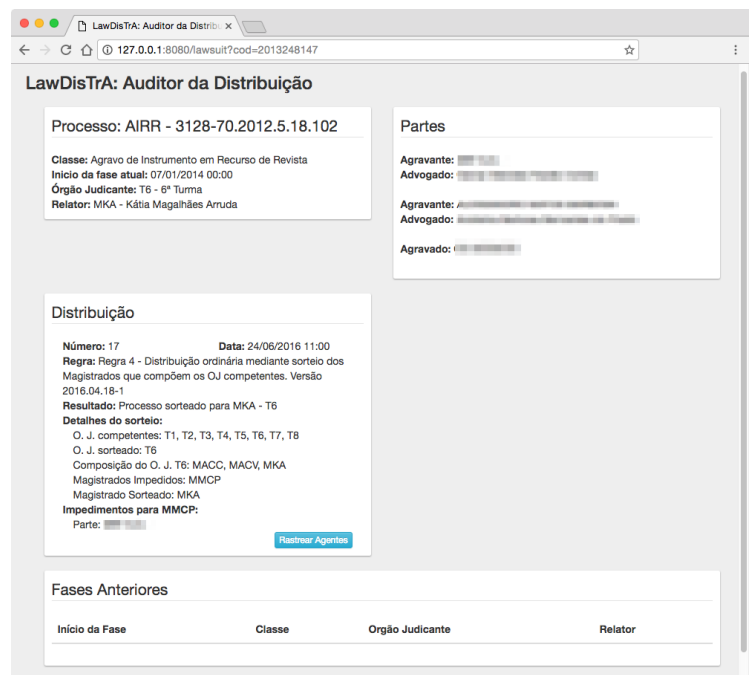

Figure 7. Distribution Auditor interface

Figure 8. Distribution agents' tracker interface

### 4.6.2. Auditability Implementation Analysis

For each of the auditability operationalization proposed in Figure 4, we discussed how they were implemented. Figure 9 presents a summary of this discussion that was organized in a table for each characteristic. Traceability is highlighted in the front as an example, according to Figure 5.

The symbols ✓, ± and • were used to represent, respectfully, high attendance, partial attendance, and no attendance.

Figure 9. Implementation of Auditability operationalization

See examples in Section 3.3 for operationalization "Each agent must register date, the actor involved, location, process status, and justification of each information change it is responsible for." The data model was designed to support the specified data. Figure 10 presents an interface that queries the distribution steps.

Section 3.3 also exemplifies the operationalization "The agents must be aware of and register the agents (and inputs) that activate it and the agents (and outputs) it activates." The schematization demanded is implemented and visualized through LawDistrA Architecture arrows (Figure 5).
In Section 4, LawDistrA is evaluated within the Brazilian lawsuit distribution at the TST.

## 5. LAWDISTRA ILLUSTRATION: A BRAZILIAN SCENARIO

LawDisTrA was experimented using information from a total of 309,332 electronic lawsuits, which is slightly more than the total number of cases received by the TST in one year[11]. 55 agents were activated: (i) 25 PAs representing 24 Regional Labor Courts and the TST; (ii) 27 MAs representing each one of the TST Magistrates; (iii) one centralized DA to perform distribution; (iv) one DF; and (v) one AMS.

LawDisTrA was able to distribute the total number of processes in 28 hours, 50 minutes and 18 seconds of uninterrupted work in a rate of 3.15 lawsuits per second using a 2011 MacBook Pro, Intel Core i7 (2,7 GHz), and 16GB DDR3 1600MHz RAM. As expected, 94.80% of lawsuits were distributed using Rule 4 (Table 1) since this rule applies to lawsuits that have no relation to each other and no relation with the Magistrates. Therefore, it is the general rule.

Table 1. Lawsuit distribution by rules

| Rule | Description | Num. Lawsuits | Frequency |
|---|---|---|---|
| 1 | Distribution by dependency of lawsuits that are related to other lawsuits already distributed | 475 | 0.15% |
| 2 | Distribution by prevention of existing lawsuits that are in new phases and must be distributed to the same Judicial Body and Magistrate of the previous phase | 15,580 | 5.04% |
| 3 | Distribution of embargoes due to divergent decisions among Judicial Bodies | 44 | 0.01% |
| 4 | Ordinary distribution by drawing of Judicial Bodies and Magistrates, which is the general distribution | 283,233 | 94.80% |

Let us take, for instance, the distribution of the Lawsuit 3128-70.2012.5.18.102 (the same process showed in Figure 7 and Figure 8), as briefly illustrated in Figure 10. This lawsuit was handled by agent PA18, which represents the Regional Court of the 18th Jurisdictional Region. PA18 checks the electronic processes database to get information, like parties interested, other related processes, previous phases, and others. PA18 (action #1) informed the distributor DA about this process waiting for distribution. DA constructs its knowledge base with metadata informed by PA18. Now DA knows that this specific lawsuit was classified as processual class AIRR (#2.1), has no relation with others (#2.2) and that it is its first distribution (#2.3). By consulting the distribution database, DA knows that processes of processual class AIRR should be distributed to one of the following JBs: T1, T2, T3, T4, T5, T6, T7 or T8 and put that on the Working Memory (WM) (#2.4).

The composition of these JBs is known by the DA because, when started, DA asked this to every MA available in the system (and asks that to every MA activated after that). Therefore, DA knows which magistrates are competent to receive the case (#2.5). Then, DA asks each MA if they have any impediment to deal with the case (#3.1). Each MA checks the parties involved in the lawsuit and answers the DA (#3.2 and 3.3). In this case, only the magistrate MMCP informed an impediment (action #3.3), so DA put that in its WM (#4). Having all information needed to carry out the distribution, DA fires the rules with the inference engine (#5). In this case, DA's perceived facts match the general distribution rule (number 4). This rule performs a drawing among the JBs, and then another drawing among the JB's MAs. As a result (#6), JB T6 was selected with MKA as Judge-Magistrate. Lastly, DA informs PA18 the results of the distribution (#7), so it can update the Electronic Processes Database and move the lawsuit to the department to be analysed.

---
[11] TST lawsuits processing: http://www.tst.jus.br/tribunal-superior-do-trabalho1

The experiment recorded 27,574,903 logs of perceptions and actions of agents. This is information that is publicized by the Distribution Auditor to enable the tracking of agents for each of the distributions performed. Through experimentation, it was perceived that LawDisTrA held distribution processes as desired, with results similar to those found in the TST, but complying with clear transparency requirements.

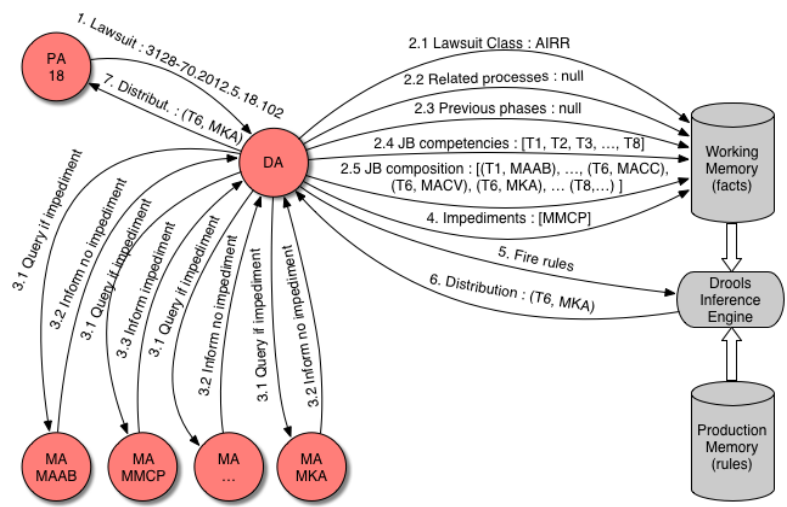

Figure 10. Example of the Lawsuit 3128-70.2012.5.18.102 distribution

## 6. DISCUSSION

This paper aimed to contribute towards an approach to systematize the auditability requirements specification of agent-based systems using Tropos methodology. LawDistrA, developed through an agent-based approach, was tested using real data from the Brazilian TST.

We noticed that the agent-based requirements specification for the lawsuit distribution system, focusing on auditability, served to demonstrate a way to define rationally and objectively auditability transparency characteristics as systems' requirements. We believe this research can also be applied considering the other four main transparency characteristics (i.e., usability, accessibility, informativeness, and understandability).

LawDistrA was evolved in two ways: (i) We systematized the formalization of the analysis of auditability characteristics during the requirements specification modeling (our main objective). Although there is still room for more software formalization, we consider this is an important initial step towards a formal interpretation of such abstract and domain-dependent concepts; and (ii) as another benefit we bring to Brazilian government is that its agent-oriented profile goes in line with the organization's performance concerning the personification of each "real" agent of the organization and also to evolve accordingly the administrative structure that may change for a variety of reasons, more usually the political ones. The architecture conceptualization facilitates the interface with other applications, which is a major concern in the Brazilian Government since public organizations, unfortunately, tend to develop their systems, with little concern of interoperability requirements.

Figure 11 presents the current interface at the TST information system, highlighting in red only the lawsuit process status "Distributed by drawing to Magistrate ACV – T6 on 03/02/2014." Note that there is much more information that contextualizes the lawsuit (as detailed in LawDistrA Implementation section) that was provided, as presented in Figures 7 and 8.

The solution shows that it provides process and information auditability to users (citizens, magistrates, organs, and any interested party) and also to the software that supports its execution. Moreover, it might provide ICT teams with a better system and process understanding for an easy evolution.

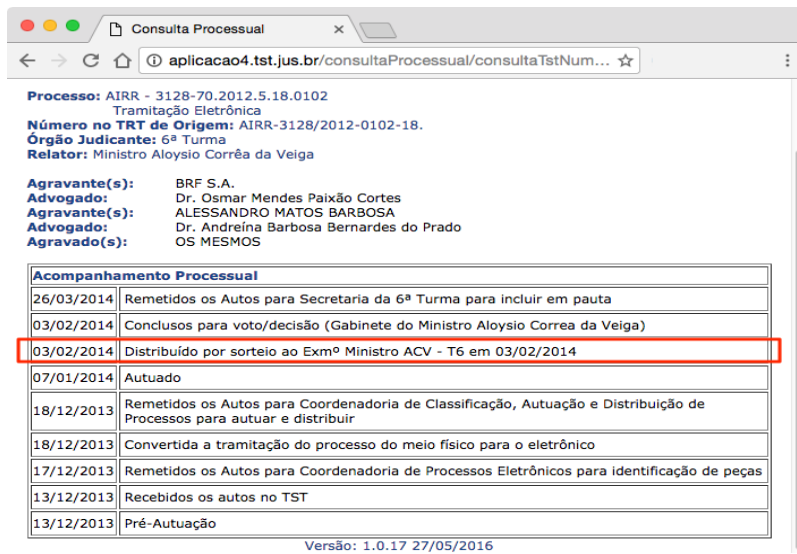

Figure 11. TST Lawsuit Consultation System (highlighted distribution information)

Limitations regarding the scenario include the fact that we conducted an empirical evaluation considering two out of five characteristics proposed in the Transparency SIG. Thus, there is room to improve with a more formal requirement elicitation. The evaluation can also be more formal, but we consider our work an important initial step towards a systematic interpretation of such abstract and domain dependent concepts.

## 7. CONCLUSIONS

This paper addresses the problem of how to systematically specify auditability requirements during agent-based system design and development based on a goal-oriented perspective. We illustrated our approach by conducting LawDisTrA design, architecture, and implementation to improve transparency in the lawsuits distribution, focusing on auditability, as described in Figures 4 and 5. The design and development of LawDisTrA using Tropos methodology enabled a better understanding of how transparency can be introduced in a traditional corporate information system using the agent paradigm. Tropos proved to be interesting to document and communicate requirements related to transparency characteristics. Besides, the agent paradigm favours the decomposition of the problem because it naturally represents how society and organizations are structured and interact.

As future work, we intend to deeply investigate the insertion of transparency requirements in business processes performed by agent-based systems. We also intend to investigate how the interaction among agents might help treat other transparency characteristics of the Transparency SIG, such as accessibility. We might perform study cases with the software development team of the TST and other organizations to further evaluate our approach in real environments.

## ACKNOWLEDGEMENTS

This research was partially supported by CAPES (grant number 1459829). Prof. Célia Ghedini Ralha thanks the support from the Brazilian National Council for Scientific and Technological Development (CNPq) for the research productivity grant number 311301/2018-5.

## REFERENCES

[1]   Fung, A., Graham, M., Weil, D. (2007) Full disclosure, the perils and promise of transparency, Cambridge University Press.

[2]   Leite, J.C.S.P., Cappelli, C. (2010) Software Transparency. Business & Information Systems Engineering, 2(3), 127-139.

[3]   Mylopoulos, J., Chung, L., Yu, E. (1999) From object-oriented to goal-oriented requirements analysis. Communications of the ACM, 42(1), 31-37.

[4]   Bresciani, P., Perini, A., Giorgini, P., Giunchiglia, F., Mylopoulos, J. (2004) Tropos: An agent-oriented software development methodology. Autonomous Agents and Multi-Agent Systems, 8(3), 203-236.

[5]   Albuquerque, D.J., Nunes, V.T., Ralha, C.G., Cappelli, C. (2016) Implementing E-government Processes Distribution with Transparency using Multi-Agent Systems. iSys: Revista Brasileira de Sistemas de Informação, 9(1), 1-21.

[6]   Albuquerque, D.J., Nunes, V.T., Ralha, C.G., Cappelli, C. (2017) E-gov Transparency Implementation Using Multi-agent System: a Brazilian Study-Case in Lawsuit Distribution Process, In Proc. of the 50th Hawaii Intern. Conf. on Systems Science (HICSS), 2772-2781.

[7]   Chung, L., Nixon, B.A., Yu, E., Mylopoulos, J. (2000) Non-functional requirements in software engineering, Springer US.

[8]   Wooldridge, M. (2009) An Introduction to MultiAgent Systems. (2nd ed.), Wiley Publishing.

[9]   Abar, S., Theodoropoulos, G.K., Lemarinier, P., O'Hare, G.M.P. (2017) Agent Based Modelling and Simulation tools: A review of the state-of-art software. Comp. Science Review, 24, 13-33.

[10]  Bellifemine, F.L., Care, G., Greenwood, D. (2007) Developing Multi-Agent Systems with JADE, Wiley Series in Agent Technology, John Wiley & Sons, Inc., USA.

[11]  Alfawzan, A., Bellamy, A. (2019) A Grounded Theory of the Requirements Engineering Process, Int. Journal of Software Engineering & Applications (IJSEA), v.10, n.5.

[12]  Yu, E., Giorgini, P., Maiden, N., Mylopoulos, J., Fickas, S. (2011) Social modeling for requirements engineering: An introduction, MIT Press.

[13]  Bertot, J.C., Jaeger, P.T., Grimes, J.M. (2010). Using ICTs to create a culture of transparency: E-government and social media as openness and anti-corruption tool for societies. Government Information Quarterly, 27(3), 264-271.

International Journal of Software Engineering & Applications (IJSEA), Vol.11, No.3, May 2020

[14] Ambekar S., Kapoor, R., Mehta, P. (2015) Structural mapping of public distribution system using multi-agent systems. Business Process Management Journal, 21(5), 1066-1090.

[15] Carneiro, D., Novais, P., Andrade, F., Zeleznikow, J., Neves, J. (2013) Using Case-based Reasoning and principled negotiation to provide decision support for dispute resolution. Knowledge and Information Systems, 36(3), 789-826.

[16] Müller, B., Balbi, S., Buchmann, C.M., Sousa, L., Dressler, G., Groeneveld, J., Klassert, C.J., Le, Q.B., Millington, J.D.A., Nolzen, H., Parker, D.C., Polhill, J.G., Schluter, M., Schulze, J., Schwarz, N., Sun, Z., Taillandier, P., Weise, H. (2014) Standardised and transparent model descriptions for agent-based models: Current status and prospects. Environmental Modelling & Software, 55, 156-163.

[17] Serrano, M., Leite, J.C.S.P. (2011) Capturing Transparency-Related Requirements Patterns through Argumentation, First International Workshop on Requirements Patterns (RePa), 32-41.

[18] Hosseini, M., Shahri, A., Phalp, K., Ali, R. (2018) Four reference models for transparency requirements in information systems. Requirements Engineering, 23(2), 251-275.

[19] Marques, J., Yelisetty, S., (2019) As Analysis of Software Requirements Specification Characteristics in Regulated Environments, Int. Journal of Software Engineering & Applications (IJSEA), v.10, n.6.

[20] Freitas, C.S., Medeiros, J.J. (2015) Organizational Impacts of the Electronic Processing System of the Brazilian Superior Court of Justice. Journal of Information Systems and Technology Management (JISTEM), 12(2), 317-332.

**AUTHORS**

Denis J. S. de Albuquerque holds a M.Sc. in Systems and Computing by the Federal University of Rio Grande do Norte (UFRN), an undergraduate's degree in Computer Science from the University of Brasília (UnB), and a Technician degree in Telecommunications by the Federal Center of Technological Education of Paraíba. Currently he is a system analyst at the 21th Regional Court of Labour, Natal, Brazil. He has experience in computer science with emphasis on databases and distributed systems, and previously worked as data administrator at the Brazilian Superior Court of Labour, Brasilia, Brazil.

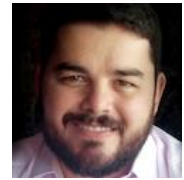

Vanessa T. Nunes holds a Ph.D. in Systems Engineering and a M.SC in Informatics from the Federal University of Rio de Janeiro (UFRJ), Brazil. She works as a collaborator researcher at the University of Brasilia and is the one of the directors of SE7Ti enterprise that focuses on consultancy and knowledge transfer of projects in the area of enterprise architecture, software development and business process management. Her current research interests include transparent computing, enterprise architecture, context management, process-aware information systems and agent-based planning.

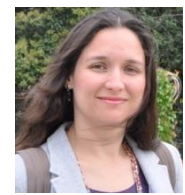

Claudia Cappelli holds a Ph.D. in Informatics from PUC-Rio and a M.SC. in Information Systems from the Federal University of Rio de Janeiro (UFRJ), Brazil, where currently she is a post-doc researcher. She works in innovation projects at Caixa Econômica Federal, Brazil. She is a researcher in transparency area at the National Institute of Science and Technology in Digital Democracy (INCT-DD), Brazil. Her current research interests include business process management, corporate architecture, information technology management, transparency and digital government.

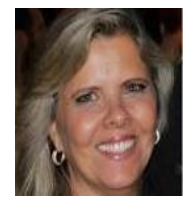

Celia G. Ralha holds a Ph.D. in Computer Science from Leeds University, England and a M.Sc. in Electronic and Computer Engineering from Aeronautics Institute of Technology (ITA), Brazil. She is an associate professor at the Department of Computer Science, University of Brasilia, Brazil. She is a senior member of the Brazilian Computer Society and receives a research productivity grant from the Brazilian National Council for Scientific and Technological Development (CNPq). Her current research interests include knowledge-based systems, multi-agent systems, agent-based modeling and simulation, and multi-agent planning.

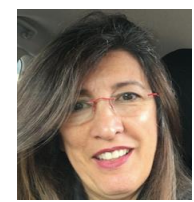